\documentclass[10pt]{sigplanconf}

\pdfoutput=1

\usepackage{amsmath}
\usepackage{graphicx}
\usepackage{url}

\usepackage{proving}

\newtheorem{criterion}{Criterion}
\newtheorem{example}{Example}

\begin{document}

\title{BliStrTune:\\ Hierarchical Invention of Theorem Proving Strategies}

\authorinfo{Jan Jakub\r{u}v\and Josef Urban}
           {Czech Technical University, CIIRC, Prague}
           {\{jakubuv,josef.urban\}@gmail.com}

\maketitle

\begin{abstract}
Inventing targeted proof search strategies for specific problem sets is a
difficult task.  State-of-the-art automated theorem provers (ATPs) such as E
allow a large number of user-specified proof search strategies described in a
rich domain specific language. Several machine learning methods that invent
strategies automatically for ATPs were proposed previously.  One of them is the
Blind Strategymaker (BliStr), a system for automated invention of ATP
strategies.

In this paper we introduce BliStrTune -- a hierarchical extension of BliStr.
BliStrTune allows exploring much larger space of E strategies by interleaving
search for high-level parameters with their fine-tuning.  We use BliStrTune to
invent new strategies based also on new clause weight functions targeted at
problems from large ITP libraries.  We show that the new strategies
significantly improve E's performance in solving problems from the Mizar
Mathematical Library.
\end{abstract}



\keywords
Automated Theorem Proving, Machine Learning, Proof Search Heuristics, Clause
Weight Functions

\section{Introduction: ATP Strategy Invention}
\label{sec:intro}

State-of-the-art automated theorem provers (ATPs) such as
E~\cite{schulz2002brainiac,Schulz13} and Vampire~\cite{Vampire} achieve their
performance by using sophisticated proof search strategies and their
combinations. Constructing good ATP search strategies is a hard task that
is potentially very rewarding. Until recently, there has been,
however, little research in this direction in the ATP community.

With the arrival of large ATP problem sets and benchmarks extracted
from the libraries of today's interactive theorem prover (ITP) systems~\cite{hammers4qed,BlanchetteGKKU16,hh4h4,holyhammer,KaliszykU13b}, automated generation of
targeted ATP strategies became an attractive topic. It seems unlikely
that manual (``theory-driven'') construction of targeted strategies
can scale to large numbers of ATP problems spanning many different
areas of mathematics and computer science.  Starting with Blind
Strategymaker (BliStr)~\cite{blistr} that was used to
invent E's strategies for MaLARea~\cite{US+08-long,malar14} on the 2012
Mizar@Turing competition problems~\cite{sutcliffe2013-j6}, several
systems have been recently developed to invent targeted ATP
strategies~\cite{SchaferS15,KuhlweinU15}. The underlying methods used
so far include genetic algorithms and iterated local search, as
popularized by the ParamILS~\cite{ParamILS-JAIR} system.

A particular problem of the methods based on iterated local search is
that their performance degrades as the number of possible strategy
parameters gets high. This is the case for E, where a domain specific
language allows construction of astronomic numbers of strategies. This
gets worse as more and more sophisticated templates for strategies are
added to E, such as our recent family of conjecture-oriented weight
functions implementing various notions of term-based
similarity~\cite{DBLP:conf/mkm/JakubuvU16}. The pragmatic solution
used in the original BliStr consisted of re-using manually
pre-designed high-level strategy components, rather than allowing the
system to explore the space of all possible strategies. This is obviously unsatisfactory.

In this work we introduce BliStrTune -- a hierarchical extension of
BliStr.  BliStrTune allows exploring much larger space of E strategies
by factoring the search into invention of good high-level strategy
components and their low-level fine-tuning.  The high-level and
low-level inventions communicate to each other their best solutions,
iteratively improving all parts of the strategy space. Together with
our new conjecture-oriented weight functions, the hierarchical
invention produces so far the strongest schedule of strategies on the
small (\emph{bushy}) versions of the Mizar@Turing problems. The
improvement over Vampire 4.0 on the training set is nearly 10\%, while
the improvement on the testing (competition) set is over 5\%.

The rest of the paper is organized as
follows. Section~\ref{sec:strats} introduces the notion of proof
search strategies, focusing on resolution/superposition ATPs and E
prover. We also summarize our recent conjecture-oriented strategies
that motivated the work on BliStrTune. Section~\ref{sec:blistr}
describes the ideas behind the original Blind Strategymaker based on
the ParamILS system (see Section~\ref{sec:paramils} for more details on
ParamILS). Section~\ref{sec:blistrtune} introduces the
hierarchical invention algorithm and its implementation. The system is
evaluated in several ways in Section~\ref{sec:experiments}, showing
significant improvements over the original BliStr and producing
significantly improved ATP strategies.

\section{Proof Search Strategies}
\label{sec:strats}
In this section we briefly describe the proof search of saturation-based automated
theorem provers (ATPs).
Section~\ref{sec:eprover} describes the proof search control possibilities of E prover~\cite{schulz2002brainiac,Schulz13}.
Section~\ref{sec:weights} describes our previous development of similarity based
clause selection strategies~\cite{DBLP:conf/mkm/JakubuvU16} which we make use of
and evaluate here.

Many state-of-the-art ATPs are based on the \emph{given clause algorithm}
introduced by \emph{Otter}~\cite{mccune1989otter,mccune1990otter,mccune1994otter}.
The input problem $T\cup\{\lnot C\}$ is translated into a refutationally equivalent
set of clauses.
Then the search for a contradiction, represented by the empty clause, is
performed maintaining two sets: the set $P$ of \emph{processed clauses} and the set
$U$ of \emph{unprocessed} clauses.
Initially, all the input clauses are unprocessed.
The algorithm repeatedly selects a \emph{given clause} $g$ from $U$ and
generates all possible inferences using $g$ and the processed clauses from $P$.
Then, $g$ is moved to $P$, and $U$ is extended with the newly produced
clauses.
This process continues until a resource limit is reached, or the empty clause is inferred, or $P$ becomes
\emph{saturated}, that is, nothing new can be inferred.

\subsection{Proof Search Strategies in E Prover}
\label{sec:eprover}

E~\cite{schulz2002brainiac,Schulz13} is a state-of-the-art theorem prover which we
use as a basis for implementation.
%
The selection of a given clause in E is implemented by a combination of
priority and weight functions.
A \emph{priority function} assigns an integer to a clause and is used to
pre-order clauses for weight evaluation.
A \emph{weight function} takes additional specific arguments
and assigns to each clause a real number called \emph{weight}.
A \emph{clause evaluation function} ($\CEF$) is specified
by a priority function, weight function, and its arguments.
Each $\CEF$ selects the clause with the smallest pair
$(\textit{priority}, \textit{weight})$ for inferences.
Each $\CEF$ is specified 
using the syntax
\[
   \heur{{WeightFunction}}{{PriorityFunction,\ldots}}
\]
with a variable number of comma separated arguments of the weight function.
E allows a user to select an \emph{expert heuristic} on a command line
in the format
\[
   \texttt{($n_1$*$\CEF_1$,\ldots,$n_k$*$\CEF_k$)}
\]
where integer $n_i$ indicates how often the corresponding 
$\CEF_i$ should be
used to select the given clause. 
E additionally supports an \emph{auto-schedule} mode where several expert
heuristics are tried, each for a selected time period.
The heuristics and time periods are automatically chosen based on 
input problem properties.

One of the well-performing weight functions in E, which we also use as a reference for
evaluation of our weight functions, is the \emph{conjecture symbol weight}.
This weight function counts symbol occurrences with different weights based
on their appearance in the conjecture as follows.
Different weights $\COEFfunc$, $\COEFconst$, $\COEFpred$,
and $\COEFvar$ are assigned 
to function, constant,
and predicate symbols, and to variables.
The weight of a symbol which appears in the conjecture is multiplied by
$\COEFconj$, typically $\COEFconj<1$ to prefer clauses with conjecture
symbols.
To compute a term weight, the given symbol weights are summed for all symbol
occurrences.
This evaluation is extended to equations and to clauses.


Apart from clause selection, E prover introduces other parameters
which influence the choice of the inference rules, term orderings, literal
selection, etc.  The selected values of the parameters which control the proof
search are called a \emph{protocol}.
Because \emph{protocol} is a crucial notion in this paper, we provide a simple
example for reader's convenience.
\begin{example}
\label{ex:proto}
Let us consider the following simplified E protocol written
in E prover command line syntax as follows.
\small
\begin{verbatim}
 -tKBO6 -WSelectComplexG 
 -H'(13*Refinedweight(PreferGoals,1,2,2,3,2), 
     2*Clauseweight(ByCreationDate,-2,-1,0.5))'
\end{verbatim}
\normalsize
This protocol selects term ordering \textup{\texttt{KBO6}}, literal selection function
\textup{\texttt{SelectComplexG}}, and two CEFs.
The first CEF has frequency 13, weight
function \textup{\texttt{Refinedweight}}, priority function
\textup{\texttt{PreferGoals}}, and
weight function arguments ``\textup{\texttt{1,2,2,3,2}}''.
An exact meaning of specific protocol parameters can be found in E
manual~\cite{Schulz13}.
\end{example}

\subsection{Similarity Based Clause Selection Strategies}
\label{sec:weights}

Many of the best-performing weight functions in E are based on a similarity of a
clause with the conjecture, for example, the \emph{conjecture symbol weight}
from the previous section.
A natural question arises whether or not it makes sense to
extend the symbol-based similarity to more complex term-based similarities.
Previously we proposed~\cite{DBLP:conf/mkm/JakubuvU16}, implemented, and
evaluated several weight functions which utilize conjecture similarity in
different ways. 
Typically they extend the symbol-based similarity by similarity on terms. Using
finer formula features improves the high-level premise selection
task~\cite{ckjujv-ijcai15}, which motivated us on steering also
the internal selection in E.
The following sections summarizes the new weight functions which we further
evaluate later in Section~\ref{sec:evaltune} and Section~\ref{sec:selectsched}.

\subsubsection{Conjecture Subterm Weight (\Term)}

The first of our weight functions  is similar to the standard \emph{conjecture symbol
weight}, counting instead of symbols the number of subterms a term shares with the conjecture.
The clause weight function \Term takes
five specific arguments $\COEFconj$, $\COEFfunc$, $\COEFconst$, $\COEFpred$
and $\COEFvar$.
The weight of a term equals weight $\COEFfunc$ for functional terms,
$\COEFconst$ for constants, $\COEFpred$ for predicates, and $\COEFvar$ for
variables, possibly multiplied by $\COEFconj$ when $\termA$ appears in the
conjecture.
To compute a clause weight, terms weights are summed for all subterms from a
clause.



\subsubsection{Conjecture Frequency Weight (\Tfidf)}

\emph{Term frequency -- inverse document frequency}, is a numerical
statistic intended to reflect how important a word is to a document
in a corpus~\cite{DBLP:books/cu/LeskovecRU14}.
A \emph{term frequency} is the number of occurrences of the term in a given
document.
A \emph{document frequency} is the number of documents in a corpus
which contain the term.
The term frequency is typically multiplied by the logarithm of the inverse of
document frequency to reduce frequency of terms which appear often.
We define $\tf{\termA}$ as 
the number of occurrences of $\termA$ in a conjecture.
We consider a fixed set of clauses denoted $\documents$.
We define $\df{\termA}$ as the count of clauses from $\documents$ which
contain $\termA$.
Out weight function \Tfidf takes one specific argument 
$\COEFdoc$ to select documents, either (1) $\coef{ax}$ for the
axioms (including the conjecture) or (2) $\coef{pro}$ for all the processed clauses.
First we define the value $\tfidf{\termA}$ of term $\termA$ as follows.
\small
\[
   \tfidf{\termA} = \tf{\termA}*\log{\frac{1+|\documents|}{1+\df{\termA}}}
\]
\normalsize
The weight of term $\termA$ is computed as 
   $\frac{1}{1+\tfidf{\termA}}$
and extended to clauses.


\subsubsection{Conjecture Term Prefix Weight (\Pref)}

The previous weight functions rely on an exact match of a term with a conjecture
related term.
The following weight function loosen this restriction and consider also
partial matches.
We consider terms as symbol sequences.
Let $\pref{\termA}$ be the longest prefix $\termA$ shares with a conjecture
term.
A \emph{term prefix weight} (\Pref) counts the length of $\pref{\termA}$
using weight arguments $\COEFmatch$ and $\COEFmiss$.
These are used to define the weight of term $\termA$ as follows.
\[
   \COEFmatch*|\pref{\termA}| + \COEFmiss*(|\termA|-|\pref{\termA}|)
\]




\subsubsection{Conjecture Levenshtein Distance Weight (\Lev)}

A straightforward extension of \Pref is to employ the Levenshtein
distance~\cite{levenshtein1966bcc} which measures a distance of two strings
as the minimum number of edit operations (character insertion, deletion,
or change) required to change one word into the other.
Our weight function \Lev defines the weight of term $\termA$ as the minimal
Levenshtein distance from $\termA$ to some conjecture term.
It takes additional arguments $\COEFins$, $\COEFdel$, $\COEFch$ to assign
different costs for edit operations.


\subsubsection{Conjecture Tree Distance Weight (\Ted)}

The Levenshtein distance does not respect a tree structure of terms.  
To achieve that, we implement the \emph{Tree edit distance} 
\cite{Zhang:1989:SFA:76071.76082} which is similar to Levenshtein but
uses tree editing operations (inserting a node into a tree, deleting a node
while reconnecting its child nodes to the deleted position, and renaming a
node label).
Our weight function \Ted takes the same arguments
as \Lev above and term weight is defined similarly.

%
%

\subsubsection{Conjecture Structural Distance Weight (\Struc)}

With \Ted, a tree produced by the edit operations does not need to represent
a valid term as the operations can change number of child nodes.
To avoid this we define a simple \emph{structural distance} which
measures a distance of two terms by a number of \emph{generalization} and
\emph{instantiation} operations.
Generalization transforms an arbitrary term to a variable while
instantiation does the reverse.
Our weight function \Struc takes additional
arguments $\COEFmiss$, $\COEFinst$, and $\COEFgen$ as penalties for
variable mismatch and operation costs.
The distance of a variable $x$ to a term $\termA$ is the cost of instantiating
$x$ by $\termA$, computed as 
      $\struc{x}{\termA} = \COEFinst*|\termA|$.
The distance of $\termA$ to $x$ is defined similarly but with $\COEFgen$.
A distance of non-variable terms $\termA$ and $\termB$ which share the
top-level symbol is the sum of distances of the corresponding arguments.
Otherwise, a generic formula 
$\struc{\termA}{x_0}+\struc{x_0}{\termB}$
is used.
The term weight is as for \Lev but using $\DSTRUC$.

%


%

\section{Blind Strategymaker (BliStr)}
\label{sec:blistr}

In this section we describe Blind Strategymaker
(BliStr)~\cite{blistr} which we further extend in the following
section.
BliStr is a system that develops E prover protocols targeted for a given large
set of problems.
The main idea is to interleave (i) iterated low-timelimit  local search for
new protocols on small sets of  similar  easy  problems  with  (ii)
higher-timelimit  evaluation  of  the new protocols  on  all  problems.
The accumulated results of the global higher-timelimit runs are used to define
and evolve the notion of  “similar  easy  problems”,  and  to  control  the
selection  of  the  next  protocol to  be  improved.

The main criterion for BliStr is as follows. 
\begin{criterion}[Max]
Invent a set of E protocols that together solve as many of the given benchmark problems.
\end{criterion}
To ensure that the invented protocols perform well also on unknown but related
problems a second criterion is considered.
\begin{criterion}[Gen]
The protocols should be reasonably general.
\end{criterion}
To simplify employment of the invented protocols, BliStr tries to achieve also
the third criterion.
\begin{criterion}[Size]
The set of such protocols should not be too large.
\end{criterion}

As defined earlier, E protocols consist of many parameters and their values which
influence the proof search.
A huge number of weight function arguments within clause evaluation
functions (CEFs, see Section~\ref{sec:eprover}) makes the set of meaningful protocol
parameters very large for a straightforward use of iterative local search as done by the ParamILS~\cite{ParamILS-JAIR} system.
Since ParamILS otherwise looks like the right tool for the task, a 
data-driven
(``blind'') approach was applied in the original BliStr to get a smaller set of meaningful CEFs:
the existing E protocols that were most useful on benchmarks of interest were
used to extract a smaller set (a dozen) of CEFs.
Making this CEFs choice more ``blind'' is the main contribution of this work
and it is discussed in details in
Section~\ref{sec:blistrtune}.


Even after such reduction, the space of the protocol parameter-value combinations is so large that a random exploration seems unlikely to find good new protocols.
The guiding idea in BliStr is to use again a data-driven approach. 
Problems in a given mathematical field often share a lot of structure and
solution methods. 
Mathematicians become better and better by solving the problems, they become
capable of doing larger and larger steps with confidence, and as a result they
can gradually attack problems that were previously too hard for them. 
By this analogy, it is plausible to think that if the solvable problems
become much easier for an ATP system, the system will be able to solve
some more (harder, but related) problems. For this to work, a method
that can improve an ATP on a set of solvable problems is needed. 
As already mentioned, the
established ParamILS system can be used for this.

\subsection{ParamILS and Its Use in the BliStr Loop}
\label{sec:paramils}
Let $A$ be an algorithm whose parameters come from a
\textit{configuration space} (product of possible values) $\Theta$.  A
\textit{parameter configuration} is an element $\theta \in \Theta$,
and $A(\theta)$ denotes the algorithm $A$ with the parameter
configuration $\theta$.  Given a distribution (set) of problem
instances $D$, the \textit{algorithm configuration problem} is to find
the parameter configuration $\theta \in \Theta$ resulting in the best
performance of $A(\theta)$ on the distribution $D$.  ParamILS is an a
implementation of an \textit{iterated local search} (ILS) algorithm
for the algorithm configuration problem. In short, starting with an
initial configuration $\theta_0$, ParamILS loops between two steps: (i) perturbing the
configuration to escape from a local optimum, and (ii) iterative 
improvement of the perturbed configuration. The result of step (ii)
is accepted if it improves the previous best configuration. 

To fully determine how to use ParamILS in a particular case, $A$,
$\Theta$, $\theta_0$, $D$, and a performance metric need to be
instantiated. In our case, $A$ is E run with a low timelimit $t_{\textit{cutoff}}$, $\Theta$
is the set of expressible E protocols, and as a performance
metric we use the number of given-clause loops done by E during solving the
problem. If E cannot solve a problem within the low
timelimit, a sufficiently high value ($10^6$) is used. 
Since it is unlikely that there is one best E protocol for all
of the given benchmark problems, 
%
it would be counterproductive to use all problems as the set $D$ for
ParamILS runs. 
Instead, BliStr partitions the set of all solvable problems
into subsets on which the particular protocols perform
best. 
See~\cite{blistr} for the technical details of the BliStr heuristic for
choosing the successive $\theta_0$ and $D$.  The complete BliStr loop
then iteratively co-evolves the set of protocols,
the set of solved problems, the matrix of the best results, and the set of
the protocols eligible for the ParamILS improvement together with their problem sets.
\\
\\

\section{BliStrTune: Hierarchical Invention}
\label{sec:blistrtune}

BliStr uses a fixed set of CEFs for inventing new protocols.  The arguments of
these fixed CEFs (the priority function, weight function arguments) cannot be
modified during the iterative protocol improvement done by ParamILS.  A
straightforward way to achieve invention (fine-tuning) of CEF arguments would be
to extend the ParamILS configuration space $\Theta$.  This, however, makes the
configuration space grow from ca. $10^7$ to $10^{120}$ of possible combinations.
Preliminary experiments revealed that with a configuration space of this size
ParamILS does not produce satisfactory results in a reasonable time.

In this section we describe our new extension of BliStr -- BliStrTune -- where
the invention of good high-level protocol parameters (Section~\ref{sec:global})
is interleaved with the invention of good CEF arguments
(Section~\ref{sec:tuning}).  The basic idea behind BliStrTune is iterated
\emph{hierarchical invention}: The large space of the optimized parameters is
naturally factored into two (in general several) layers, and at any time only
one layer is subjected to invention, while the other layer(s) remain fixed. The
results then propagate between the layers, and the layer-tuning and propagation
are iterated.  BliStrTune is experimentally evaluated in
Section~\ref{sec:experiments}.


\subsection{Global Parameter Invention}
\label{sec:global}

The ParamILS runs used in the BliStrTune's global-tuning phase are essentially the same
as in the case of BliStr, with the following minor exceptions.
BliStr uses a fixed configuration space $\Theta$ for all ParamILS runs.  This is
possible because a small set (currently 12) of CEFs is hard coded in Blistr's
$\Theta$.
BliStrTune uses in the global-tuning phase a parametrized configuration space
$\Theta_{C}$ where $C$ is a collection of CEFs that can be different for each
ParamILS run.
This collection can be arbitrary but we use only the 50 best performing
CEFs in order to limit the configuration space size for the global-tuning phase.
The notion of ``best performing CEFs'' develops in time and it is discussed in
details in Section~\ref{sec:collection}.
Furthermore, BliStrTune introduces additional argument $c_\textit{cef}$ to limit
the maximum number of CEFs which can occur in a single protocol
($c_\textit{cef}=12$ for the case of BliStr).

BliStrTune's global-tuning usage of ParamILS is otherwise the same as in BliStr,
that is, given $\Theta_{C}$, the initial configuration $\theta_0\in\Theta_{C}$,
and problems $D$, the result of the global tuning is a configuration
$\theta_1\in\Theta_{C}$ which has the best found performance on $D$.
This configuration $\theta_1$ then serves as an input for the next fine-tuning phase.

\begin{example}
\label{ex:example}
Let us consider the E protocol from Example~\ref{ex:proto}.
In the global-tuning phase we instruct ParamILS to modify top level arguments,
that is, term ordering (``\textup{\texttt{-t}}''), literal selection
(``\textup{\texttt{-W}}''), CEF
frequencies (``\textup{\texttt{13*}}'' and ``\textup{\texttt{2*}}''), and also the whole CEF
blocks and their count.
We do not, however, allow ParamILS to change CEF arguments (priority functions
and weight function arguments).
The whole CEF must be changed to another CEF from collection $C$.
\end{example}

\subsection{Invention of the CEF Arguments}
\label{sec:tuning}

Given the result of the global-tuning phase $\theta_1\in\Theta_{C}$ a new configuration space for the
fine-tuning phase $\Theta_{\theta_1}$ is constructed by (1) fixing the parameter
values from $\theta_1$ and by (2) an introduction of new parameters that allow
to change the values of the arguments of the CEFs used in $\theta_1$.
In order to do that, we need to describe the space of the possible values of the CEF arguments.

The CEF arguments (see Section~\ref{sec:eprover}) consist of the priority function and the weight
function specific arguments.
Because of the different number and semantics of the weight function arguments, we do not allow to change the CEF's weight functions during the fine-tuning. They are fixed
to the values provided in $\theta_1$. 
For each weight function argument, we know its type (such as the symbol \emph{weight},
operation \emph{cost}, weight \emph{multiplier}, etc.). For each type we have pre-designed the
set of reasonable values.
For the original E weight functions, we extract the reasonable values from the 
auto-schedule mode of E.
For our new weight functions, we use our preliminary
experiments~\cite{DBLP:conf/mkm/JakubuvU16} enhanced with our intuition.

Given the configuration space $\Theta_{\theta_1}$, a configuration
$\theta_1\in\Theta_C$ can be easily converted to an equivalent configuration
$\theta_1'\in\Theta_{\theta_1}$ by setting the parameter values to those CEFs
arguments that were previously fixed in $\theta_1$ and $C$.
Then we can run ParamILS with the configuration space $\Theta_{\theta_1}$, the
initial configuration $\theta_1'$, and with the same problem set $D$ as in the
global-tuning phase.
The result is a configuration $\theta_2'\in\Theta_{\theta_1}$ providing the best found
performance on $D$.

The global invention (global tuning) and the local invention (fine-tuning) phases can be iterated. To do that, we  
need to transform the result of the fine-tuning $\theta_2'\in\Theta_{\theta_1}$ to an
equivalent initial configuration $\theta_2\in\Theta_C$ for the next global-tuning
phase.
In order to do that, the CEFs invented by $\theta_2'$ must be present in the CEFs
collection $C$.
If this is not the case, we simply extend $C$ with the new CEFs.
In practice, we now use two iterations of this process (that is, two phases of
global-tuning and two phases of fine-tuning) which was experimentally evaluated
to provide good results.

\begin{example}
Recall the protocol from Example~\ref{ex:proto} and Example~\ref{ex:example}.
In the fine-tuning phase we would fix all the top level arguments modified in
global-tuning phase (``\textup{\texttt{-t}}'', and so on, as described in
Example~\ref{ex:example}) and we would instruct
ParamILS to change individual CEF arguments.
That is, the values
\small
\begin{verbatim}
   PreferGoals,1,2,2,3,2 
   ByCreationDate,-2,-1,0.5
\end{verbatim}
\normalsize
might be changed to different values while the rest of the protocol stays
untouched.
\end{example}

\subsection{Maintaining Collections of CEFs}
\label{sec:collection}

The global-tuning phase of BliStrTune requires the collection $C$ of CEFs as an
input.
It is desirable that this collection $C$ is limited in size (currently we use
max.~50 CEFs) and that it contains the best performing CEFs.

Initially, for each weight function $w$ defined in E, we have extracted the CEF
most often used in the E auto-schedule mode.
We have added a CEF for each of our new weight functions.
This gave us the initial collection of 21 CEFs.
Then we use a global database (shared by different BliStrTune runs) in which we
store all CEFs together with the usage counter which states how often each CEF
was used in a protocol invented by BliStrTune.
Recall that in one BliStrTune iteration, ParamILS is ran four times (two phases
of global-tuning and two phases of fine-tuning).
Whenever a CEF is contained in a protocol invented by any BliStrTune iteration
(after the four ParamILS runs), we increase the CEF usage counter, perhaps
adding a new CEF to the database when used for the first time.

To select the 50 best performing CEFs we start with $C=\emptyset$.
We extract all the weight functions $W$ used in the global CEF database.
This set $W$ stays constant because the database already contains all possible
weight functions from the very beginning.
For each $w\in W$, we compute the list $C_w$ of all CEFs from the database which
use $w$ and sort it by the usage counter.
Then we iterate over $W$ and for each $w$ we move the most often used CEF from
$C_w$ to $C$.
We repeat this until $C$ has the desirable size (or we are out CEFs).
This ensures that $C$ contains at least one CEF for each weight function.


\section{Experimental Evaluation}
\label{sec:experiments}

This section provides an experimental evaluation%
\footnote{All the experiments were run on 2x16 cores Intel(R) Xeon(R) CPU E5-2698
v3 @ 2.30GHz with 128 GB memory.  One prover run was however limited to 1 GB
memory limit.}%
of BliStrTune system.
In Section~\ref{sec:evaltune} we compare our improved BliStrTune with the
original BliStr, and we use BliStrTune to evaluate the value added by the new weight
functions.
In Section~\ref{sec:evalvar} we evaluate the BliStrTune runs with different
parameters.
In Section~\ref{sec:selectsched} we discuss and compare several methods to
construct a protocol scheduler that tries several protocols to solve a problem.
Section~\ref{sec:evalsched} then compares the best protocol scheduler with
state-of-the-art ATPs, namely, with E 1.9 using its auto-schedule mode and with Vampire
4.0.

For the evaluation we use problems from the Mizar@Turing division of the CASC 2012 (Turing100) competition mentioned in
Section~\ref{sec:intro}.
These problems come from the MPTP translation~\cite{Urb04-MPTP0,Urban06,abs-1108-3446} of the Mizar Mathematical Library~\cite{mizar-in-a-nutshell}.
The problems are divided into 1000 training and 400 testing problems.
The training problems were published before the competition, while the testing problems were
used in the competition.
This fits our evaluation setting: we can use BliStrTune to invent targeted
protocols for the training problems and then evaluate them on the testing
problems.

\subsection{Hierarchical Invention and Weight Functions}
\label{sec:evaltune}

\begin{figure}
\begin{center}
\includegraphics[width=8.5cm]{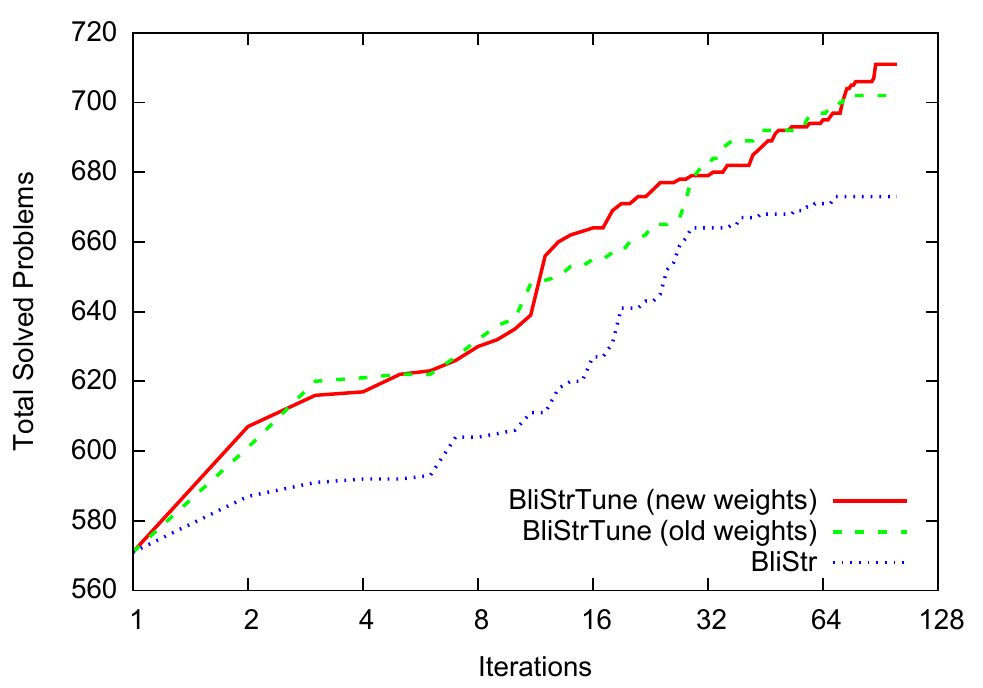}
\end{center}
\caption{Value added by parameter fine-tuning and by new weight functions
(Section~\ref{sec:evaltune}).}
\label{fig:improvement}
\end{figure}

To evaluate the hierarchical invention we ran BliStr and BliStrTune
with equivalent arguments.
Furthermore, we ran two instances of BliStrTune to evaluate the performance added by
the new weight functions from Section~\ref{sec:weights}.
The first instance was allowed to use only the original E 1.9 weight functions, while
the second additionally used our new weight functions.

BliStr and BliStrTune used the same input arguments.
The first argument is the set of the training problems. We use the 1000 training problems
from the Mizar@Turing competition in all experiments.
Other arguments are:
\begin{description}
\item[$T_\textit{improve}$] the time limit (seconds) for one ParamILS run,
\item[$t_\textit{cutoff}$] the time limit for E prover runs within ParamILS,
\item[$t_\textit{eval}$] the time limit for the protocol evaluation in BliStr/Tune.
\end{description}
In BliStrTune, ParamILS is run four times in each iteration, hence we set
$T_\textit{improve}=100$ in BliStrTune and $T_\textit{improve}=400$ in
BliStr.~\footnote{So that the times used to improve a protocol are equal.}
We set $t_\textit{cutoff}=1$ and $t_\textit{eval}=5$ and additionally, in the
case of BliStrTune, $c_\textit{cef}=6$.

The results are shown in Figure~\ref{fig:improvement}.
In each iteration (x-axis, logarithmic scale) we count the total number of the
training problems solved (y-axis) by all the protocols invented so far, provided
each protocol is given the time limit $t_\textit{eval}$.
This metric gives us relatively good idea of the BliStr/Tune progress.

The original BliStr solved 673 problems, BliStrTune without the new weights solved
702 problems, while BliStrTune with the new weights solved 711 problems.
From this and from the figure we can see that the greatest improvement is thanks to
the hierarchical parameter invention.
However, the new weight functions still provide 9 more solved problems which
is a useful additional improvement.

\subsection{Influence of the BliStrTune Input Arguments}
\label{sec:evalvar}

\begin{table*}
\begin{center}
\def\arraystretch{1.3}%
\begin{tabular}{rrrr||rrrrrr}
$T_\textrm{\textit{improve}}$ & 
$t_\textrm{\textit{cutoff}}$ & 
$t_\textrm{\textit{eval}}$ & 
$c_\textit{cef}$ & 
iters &
protos & 
run time &
best proto &
$\textbf{solved}$ &
\textit{useful}
\\\hline
100 & 1 & 5  & 6  & 115 & 116 &  1d0h & 572 & \textbf{711} & \textit{28\%} \\
100 & 1 & 5  & 10 & 111 & 115 &  1d3h & 594 & \textbf{715} & \textit{14\%} \\
300 & 1 & 5  & 6  & 83  &  87 & 1d13h & 596 & \textbf{698} & \textit{4\%} \\
300 & 1 & 5  & 10 & 82  &  85 & 1d22h & 611 & \textbf{711} & \textit{11\%} \\\hline
100 & 2 & 10 & 6  & 152 & 148 & 1d20h & 579 & \textbf{720} & \textit{27\%} \\
100 & 2 & 10 & 10 & 88  &  88 &  1d4h & 567 & \textbf{698} & \textit{1\%} \\
300 & 2 & 10 & 6  & 153 & 153 & 3d18h & 583 & \textbf{727} & \textit{19\%} \\
300 & 2 & 10 & 10 & 139 & 139 &  3d9h & 587 & \textbf{719} & \textit{15\%} \\
\end{tabular}
\end{center}
\caption{Evaluation of different BliStrTune training runs on Mizar@Turing 
problems (Section~\ref{sec:evalvar}).}
\label{tab:blistrtune}
\end{table*}

In this section we evaluate several BliStrTune runs with different input
arguments.
We run all the combinations of 
$T_\textit{improve}\in\{100,300\}$ and
$c_\textit{cef}\in\{6,10\}$ and 
$t_\textit{cutoff}\in\{1,2\}$.
This gives us 6 different BliStrTune runs.
We always set $t_\textit{eval}=5\cdot t_\textit{cutoff}$.

The results are summarized in Table~\ref{tab:blistrtune}.  Column \emph{iters}
contains the number of iterations executed by the appropriate BliStrTune run,
\emph{proto} is the total number of protocols generated, \emph{run time} is the
total run time of the given BliStrTune run, \emph{best proto} is the number of
training problems solved by the best protocol within $t_\textit{eval}$ time
limit, and \emph{solved} is the total number of the training problems solved by
all the generated protocols, provided each protocol is given time limit
$t_\textit{eval}$.
We can see that a huge amount protocols were generated.  Only few of them were
used for the final evaluation as described in Section~\ref{sec:selectsched}.
Those used for the final evaluation are considered ``useful'' and the column
\emph{useful} states how many percent of the useful protocols come from the appropriate
BliStrTune run.

We can see that the most useful runs are the basic runs with smaller
$T_\textit{improve}$ which also have lower run times.
Higher $T_\textit{improve}$ leads to higher run times but it produces better
protocols in the sense that a smaller number of protocols can solve equal
number of problems.
From the table we can see that when $t_\textit{cutoff}$ and $c_\textit{cef}$ are increased,
$T_\textit{improve}$ should be increased as well to provide ParamILS enough time
for protocol improvement.

\subsection{Selecting Best Protocol Scheduler}
\label{sec:selectsched}

\begin{table}
\begin{center}
\def\arraystretch{1.3}%
\begin{tabular}{r||r|rr|rr}
   &    & \multicolumn{2}{c|}{\textbf{training}} & \multicolumn{2}{c}{\textit{testing}} \\
scheduler & protos  & \textbf{solved} & \textbf{V+} & \textit{solved} & \textit{V+}
\\\hline
$\SidGreedy_{1}$  & 33 & \textbf{744} & \textbf{+9.8\%}    & \textit{280} & \textit{+5.2\%}  \\
$\SidGreedy_{2}$  & 27 & 742 & +9.6\%                      & \textit{279} & \textit{+4.8\%}    \\
$\SidGreedy_{5}$  & 28 & 734 & +8.4\%                      & \textit{280} & \textit{+5.2\%}    \\
$\SidGreedy_{10}$ & 22 & 719 & +6.2\%                      & \textit{276} & \textit{+3.8\%}    \\\hline
$\SidSOTAC_{15}$  & 15 & 663 & -2.0\%                      & \textit{261} & \textit{-1.8\%}    \\
$\SidSOTAC_{30}$  & 30 & 693 & +2.3\%                      & \textit{266} & \textit{+0\%}    \\
$\SidSOTAC_{45}$  & 45 & 698 & +3.1\%                      & \textit{270} & \textit{+1.5\%}    \\
$\SidSOTAC_{60}$  & 60 & \textbf{699} & \textbf{+3.2}\%                      & \textit{270} & \textit{+1.5\%}    \\\hline
$\SidESOTAC_{15}$ & 15 & 692 & +2.2\%                      & \textit{268} & \textit{+0.7\%}    \\
$\SidESOTAC_{30}$ & 30 & 711 & +5.0\%                      & \textit{273} & \textit{+2.6\%}    \\
$\SidESOTAC_{45}$ & 45 & \textbf{712} & \textbf{+5.1}\%                      & \textit{276} & \textit{+3.8\%}    \\
$\SidESOTAC_{60}$ & 60 & 707 & +4.4\%                      & \textit{275} & \textit{+3.4\%}    \\
\end{tabular}
\end{center}
\caption{BliStrTune schedulers evaluation in 60 seconds on Mizar@Turing problems (Section~\ref{sec:selectsched}).}
\label{tab:eval}
\end{table}

The 6 runs of BliStrTune described above in Section~\ref{sec:evalvar} generated
more than 900 different protocols.
In this section we try to select the best subset of protocols and construct a
\emph{protocol scheduler} which sequentially tries several protocols to solve a
problem.
We only experiment with the simplest schedulers where the time limit for solving
a problem is equally distributed among all the protocols within a scheduler.
Hence the problem of scheduler construction is reduced to the selection of the
right protocols.

We use three different ways to select scheduler protocols.
Firstly we use a \emph{greedy} approach as follows.
We evaluate all the protocols on all the training problems with a fixed time
limit $t$.
Then we construct a greedy covering sequence which starts with the best
protocol, and each next protocol in the sequence is the protocol that adds most
solutions to the union of problems solved by all previous protocols in the
sequence.
The resulting scheduler is denoted $\SidGreedy_t$.

Second way to construct a scheduler is using \emph{state-of-the-art
contribution} (SOTAC) used by CASC.
A SOTAC for the problem is the inverse of the number of protocols that solved the problem.
A protocol SOTAC is the average SOTAC over the problems it solves.
We can sort the protocols by SOTAC and select first $n$ protocols from this
sequence.
The resulting scheduler is denoted $\SidSOTAC_n$.

SOTAC of a protocol will be high even if the protocol solves only one problem
which no other protocol can solve. 
That is why also the $\SidESOTAC$ value~\cite{holyhammer} is introduced:
the sum of problem SOTAC over all the problems.
This gives us schedulers denoted $\SidESOTAC_n$.

The evaluation of 12 different schedulers with 60 seconds time limit on the
training problems is provided in Table~\ref{tab:eval}.
Column \emph{protos} specifies the count of protocols within the scheduler.
We shall use this evaluation to select the best scheduler, hence the results on
the 400 testing problems are provided for reference only.
Column \emph{solved} is the number of problems solved in 60s.
Column \emph{V+} is a percentage gain/lost on a state-of-the-art prover Vampire
4.0 which solves 667 of the 1000 training problems and 266 of the 400 testing problems.

We can see that the best results are achieved by scheduler $\SidGreedy_1$, which
also gives the best results on the testing problems.
Generally, it is better to run a bigger number of protocols with lower
individual time limit.

Furthermore, we can use the constructed schedulers to evaluate the contribution
of our new weight functions by analyzing weight functions used in the
schedulers.
Table~\ref{tab:weights} summarizes the usage of different weight functions in
the final schedulers.
Our weight functions are referred to by their names from
Section~\ref{sec:weights} while the original weights are called by their E
prover names.
Column \emph{count} states how many times the corresponding weight function was
used in some scheduler protocol, while column \emph{freq} sums the frequencies
of occurrences of CEFs which use the given weight function.
We can see that our new weight function \Term was the most often used weight
function.
Four of our weight functions were, however, not used very often which we
attribute to their higher time complexity.

\subsection{Best Protocol Scheduler Evaluation}
\label{sec:evalsched}

\begin{table}
\begin{center}
\def\arraystretch{1.3}%
\begin{tabular}{r||rr|rr}
       & \multicolumn{2}{c|}{\textit{training}} & \multicolumn{2}{c}{\textbf{testing}} \\
prover & \textit{solved} & \textit{V+} & \textbf{solved} & \textbf{V+}
\\\hline
E (BliStrTune)      & \textit{744} & \textit{+9.8\% } & \textbf{280} & \textbf{+5.2\%}  \\
Vampire 4.0         & \textit{677} & \textit{+0\%   } & \textbf{266} & \textbf{+0\%}    \\
E (auto-schedule)   & \textit{605} & \textit{-10.6\%} & \textbf{231} & \textbf{-13.1\%}
\end{tabular}
\end{center}
\caption{Evaluation of the best BliStrTune scheduler on testing problems with
60 seconds time limit.}
\label{tab:final}
\end{table}

\begin{figure}
\begin{center}
\includegraphics[width=8.5cm]{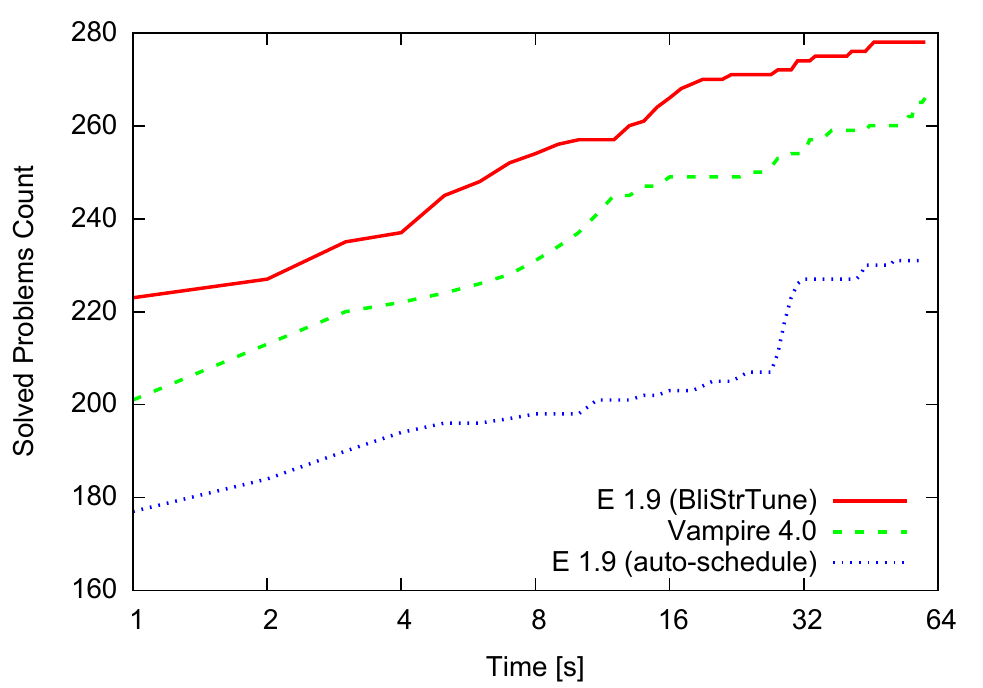}
\end{center}
\caption{Progress of ATPs on the 400 Mizar@Turing testing problems with 60 seconds time limit.}
\label{fig:progress}
\end{figure}

In this section we evaluate the best protocol scheduler $\SidGreedy_1$ selected
in previous Section~\ref{sec:selectsched} on the testing problems with 60
seconds time limit.
We compare $\SidGreedy_1$ with two state-of-the-art ATPs: (1) with E prover 1.9
in auto-schedule mode and (2) with Vampire 4.0 in CASC mode.

The results are summarized in Table~\ref{tab:final}.
We can see that E with scheduler $\SidGreedy_1$ invented by BliStrTune
outperforms Vampire by 5.2\% and the improvement from E in auto-schedule mode is
even more significant.
Figure~\ref{fig:progress} provides a graphical representation of ATP's progress.
For each second (x-axis, logarithmic scale) we count the number of problems solved so
far (y-axis).
We can see that $\SidGreedy_1$ was outperforming Vampire during the whole
evaluation.

\section{Conclusions and Future Work}

In this paper we have described BliStrTune, an extension of a previously published
system BliStr, which can be used for hierarchical invention of protocols
targeted for a given benchmark problems.
The main contribution of BliStrTune is that it considers a much bigger space of
protocols by interleaving the global-tuning phase with argument fine-tuning.
We have evaluated the original BliStr and our BliStrTune on the same input data and
experimentally proved that BliStrTune outperforms BliStr.
We have evaluated several ways of creating protocol schedulers and showed that E
1.9 with the best protocol scheduler constructed from BliStrTune protocols
targeted for training problems outperforms state-of-the-art ATP Vampire 4.0 on
independent testing problems by more than 5\%.

Furthermore, we have used BliStrTune to evaluate a contribution of our previously
designed weight functions in E prover.
We have shown that the new weight functions allow us to solve more problems and
that (at least two of them) were often used in the best scheduler protocols.
Interestingly, more complex structural weights (like \Lev, \Ted) were not used
very often in the schedulers even though our previous experiments suggested
they might be very useful.
We attribute this to their higher time complexity and we would like to
investigate this in our future research.

Several topics are suggested for future work.
We have shown that new weight functions can enhance E prover performance, hence
more weight functions which consider term structure could be implemented.
It seems that it will be better to design weight functions with lower time
complexity, perhaps even providing approximate results (for example,
some approximation of the Levenshtein distance which could be computed faster).

Another direction of our future research is to design more complex protocol schedulers.
We have achieved good results with the simplest protocol schedulers where each
protocol is given an equal amount of time when solving a problem.
It would be interesting to design ``smarter'' schedulers and to see how many
more problems can be solved.

Further direction of our future research are enhancements of the BliStr/Tune main loop.
We could experiment with settings of various parameters, or with selection of
training problems, or we could use parameter improvement
methods other than ParamILS~\cite{wang2016bayesian}.
Finally, we would like to make our implementation easier to use and to
distribute it as a solid software package.

\begin{table}
\begin{center}
\def\arraystretch{1.3}%
\begin{tabular}{r||rr}
weight function & count & freq \\\hline
\Term & \textbf{111} & \textbf{782} \\
\textrm{RelevanceLevelWeight2} & 104 & 353 \\
\Pref & \textbf{45} & \textbf{297} \\
\textrm{ConjectureGeneralSymbolWeight} & 43 & 235 \\
\textrm{FIFOWeight} & 39 & 174 \\
\textrm{StaggeredWeight} & 36 & 199 \\
\textrm{SymbolTypeweight} & 34 & 96 \\
\textrm{ConjectureRelativeSymbolWeight} & 28 & 110 \\
\textrm{ConjectureSymbolWeight} & 23 & 44 \\
\textrm{RelevanceLevelWeight} & 21 & 40 \\
\textrm{OrientLMaxWeight} & 13 & 24 \\
\textrm{Refinedweight} & 12 & 88 \\
\textrm{Defaultweight} & 10 & 131 \\
\textrm{Clauseweight} & 9 & 64 \\
\textrm{ClauseWeightAge} & 6 & 10 \\
\textrm{PNRefinedweight} & 5 & 21 \\
\Struc & \textbf{5} & \textbf{14} \\
\Lev & \textbf{3} & \textbf{37} \\
\textrm{Uniqweight} & 2 & 2 \\
\Ted & \textbf{1} & \textbf{13} \\
\Tfidf & \textbf{1} & \textbf{1} \\\hline
total & 551 & 2735 \\
\end{tabular}
\end{center}
\caption{Usage of weight functions in best schedulers.}
\label{tab:weights}
\end{table}
\appendix

\acks

Supported by the ERC Consolidator grant number 649043 \textit{AI4REASON}.


\bibliographystyle{abbrvnat}

\begin{thebibliography}{28}
\providecommand{\natexlab}[1]{#1}
\providecommand{\url}[1]{\texttt{#1}}
\expandafter\ifx\csname urlstyle\endcsname\relax
  \providecommand{\doi}[1]{doi: #1}\else
  \providecommand{\doi}{doi: \begingroup \urlstyle{rm}\Url}\fi

\bibitem[Alama et~al.(2014)Alama, Heskes, K\"{u}hlwein, Tsivtsivadze, and
  Urban]{abs-1108-3446}
J.~Alama, T.~Heskes, D.~K\"{u}hlwein, E.~Tsivtsivadze, and J.~Urban.
\newblock Premise selection for mathematics by corpus analysis and kernel
  methods.
\newblock \emph{J. Autom. Reasoning}, 52\penalty0 (2):\penalty0 191--213, 2014.
\newblock ISSN 0168-7433.
\newblock \doi{10.1007/s10817-013-9286-5}.

\bibitem[Blanchette et~al.(2016{\natexlab{a}})Blanchette, Kaliszyk, Paulson,
  and Urban]{hammers4qed}
J.~Blanchette, C.~Kaliszyk, L.~Paulson, and J.~Urban.
\newblock Hammering towards {QED}.
\newblock \emph{Journal of Formalized Reasoning}, 9\penalty0 (1):\penalty0
  101--148, 2016{\natexlab{a}}.
\newblock ISSN 1972-5787.
\newblock URL \url{http://jfr.unibo.it/article/view/4593}.

\bibitem[Blanchette et~al.(2016{\natexlab{b}})Blanchette, Greenaway, Kaliszyk,
  K{\"{u}}hlwein, and Urban]{BlanchetteGKKU16}
J.~C. Blanchette, D.~Greenaway, C.~Kaliszyk, D.~K{\"{u}}hlwein, and J.~Urban.
\newblock A learning-based fact selector for {Isabelle/HOL}.
\newblock \emph{J. Autom. Reasoning}, 57\penalty0 (3):\penalty0 219--244,
  2016{\natexlab{b}}.
\newblock \doi{10.1007/s10817-016-9362-8}.
\newblock URL \url{http://dx.doi.org/10.1007/s10817-016-9362-8}.

\bibitem[Gauthier and Kaliszyk(2015)]{hh4h4}
T.~Gauthier and C.~Kaliszyk.
\newblock Premise selection and external provers for {HOL4}.
\newblock In \emph{Certified Programs and Proofs (CPP'15)}, LNCS. Springer,
  2015.
\newblock \doi{10.1145/2676724.2693173}.
\newblock URL \url{http://dx.doi.org/10.1145/2676724.2693173}.
\newblock \url{http://dx.doi.org/10.1145/2676724.2693173}.

\bibitem[Grabowski et~al.(2010)Grabowski, Korni{\l}owicz, and
  Naumowicz]{mizar-in-a-nutshell}
A.~Grabowski, A.~Korni{\l}owicz, and A.~Naumowicz.
\newblock {M}izar in a nutshell.
\newblock \emph{J. Formalized Reasoning}, 3\penalty0 (2):\penalty0 153--245,
  2010.

\bibitem[Hutter et~al.(2009)Hutter, Hoos, Leyton-Brown, and
  St\"{u}tzle]{ParamILS-JAIR}
F.~Hutter, H.~H. Hoos, K.~Leyton-Brown, and T.~St\"{u}tzle.
\newblock {ParamILS:} an automatic algorithm configuration framework.
\newblock \emph{J. Artificial Intelligence Research}, 36:\penalty0 267--306,
  October 2009.

\bibitem[Jakub\r{u}v and Urban(2016)]{DBLP:conf/mkm/JakubuvU16}
J.~Jakub\r{u}v and J.~Urban.
\newblock Extending {E} prover with similarity based clause selection
  strategies.
\newblock In \emph{Intelligent Computer Mathematics - 9th International
  Conference, {CICM} 2016, Bialystok, Poland, July 25-29, 2016, Proceedings},
  pages 151--156, 2016.

\bibitem[Kaliszyk and Urban(2014)]{holyhammer}
C.~Kaliszyk and J.~Urban.
\newblock Learning-assisted automated reasoning with {F}lyspeck.
\newblock \emph{J. Autom. Reasoning}, 53\penalty0 (2):\penalty0 173--213, 2014.
\newblock \doi{10.1007/s10817-014-9303-3}.
\newblock URL \url{http://dx.doi.org/10.1007/s10817-014-9303-3}.

\bibitem[Kaliszyk and Urban(2015)]{KaliszykU13b}
C.~Kaliszyk and J.~Urban.
\newblock {MizAR 40 for Mizar 40}.
\newblock \emph{J. Autom. Reasoning}, 55\penalty0 (3):\penalty0 245--256, 2015.
\newblock \doi{10.1007/s10817-015-9330-8}.
\newblock URL \url{http://dx.doi.org/10.1007/s10817-015-9330-8}.

\bibitem[Kaliszyk et~al.(2015{\natexlab{a}})Kaliszyk, Urban, and
  Vyskocil]{ckjujv-ijcai15}
C.~Kaliszyk, J.~Urban, and J.~Vyskocil.
\newblock Efficient semantic features for automated reasoning over large
  theories.
\newblock In Q.~Yang and M.~Wooldridge, editors, \emph{IJCAI'15}, pages
  3084--3090. {AAAI} Press, 2015{\natexlab{a}}.

\bibitem[Kaliszyk et~al.(2015{\natexlab{b}})Kaliszyk, Urban, and
  Vyskocil]{malar14}
C.~Kaliszyk, J.~Urban, and J.~Vyskocil.
\newblock Machine learner for automated reasoning 0.4 and 0.5.
\newblock In S.~Schulz, L.~D. Moura, and B.~Konev, editors, \emph{PAAR-2014.
  4th Workshop on Practical Aspects of Automated Reasoning}, volume~31 of
  \emph{EPiC Series in Computing}, pages 60--66. EasyChair, 2015{\natexlab{b}}.

\bibitem[Kov{\'a}cs and Voronkov(2013)]{Vampire}
L.~Kov{\'a}cs and A.~Voronkov.
\newblock First-order theorem proving and {V}ampire.
\newblock In N.~Sharygina and H.~Veith, editors, \emph{CAV}, volume 8044 of
  \emph{LNCS}, pages 1--35. Springer, 2013.
\newblock ISBN 978-3-642-39798-1.

\bibitem[K{\"{u}}hlwein and Urban(2015)]{KuhlweinU15}
D.~K{\"{u}}hlwein and J.~Urban.
\newblock {MaLeS}: {A} framework for automatic tuning of automated theorem
  provers.
\newblock \emph{J. Autom. Reasoning}, 55\penalty0 (2):\penalty0 91--116, 2015.
\newblock \doi{10.1007/s10817-015-9329-1}.
\newblock URL \url{http://dx.doi.org/10.1007/s10817-015-9329-1}.

\bibitem[Leskovec et~al.(2014)Leskovec, Rajaraman, and
  Ullman]{DBLP:books/cu/LeskovecRU14}
J.~Leskovec, A.~Rajaraman, and J.~D. Ullman.
\newblock \emph{Mining of Massive Datasets, 2nd Ed}.
\newblock Cambridge University Press, 2014.
\newblock ISBN 978-1107077232.

\bibitem[Levenshtein(1966)]{levenshtein1966bcc}
V.~Levenshtein.
\newblock {Binary Codes Capable of Correcting Deletions, Insertions and
  Reversals}.
\newblock \emph{Soviet Physics Doklady}, 10:\penalty0 707, 1966.

\bibitem[McCune(1990)]{mccune1990otter}
W.~McCune.
\newblock Otter 2.0.
\newblock In \emph{International Conference on Automated Deduction}, pages
  663--664. Springer, 1990.

\bibitem[McCune(1989)]{mccune1989otter}
W.~W. McCune.
\newblock Otter 1. 0 users' guide.
\newblock Technical report, Argonne National Lab., IL (USA), 1989.

\bibitem[McCune(1994)]{mccune1994otter}
W.~W. McCune.
\newblock \emph{Otter 3.0 reference manual and guide}, volume 9700.
\newblock Argonne National Laboratory Argonne, IL, 1994.

\bibitem[Sch{\"{a}}fer and Schulz(2015)]{SchaferS15}
S.~Sch{\"{a}}fer and S.~Schulz.
\newblock Breeding theorem proving heuristics with genetic algorithms.
\newblock In G.~Gottlob, G.~Sutcliffe, and A.~Voronkov, editors, \emph{Global
  Conference on Artificial Intelligence, {GCAI} 2015, Tbilisi, Georgia, October
  16-19, 2015}, volume~36 of \emph{EPiC Series in Computing}, pages 263--274.
  EasyChair, 2015.
\newblock URL
  \url{http://www.easychair.org/publications/paper/Breeding_Theorem_Proving_Heuristics_with_Genetic_Algorithms}.

\bibitem[Schulz(2002)]{schulz2002brainiac}
S.~Schulz.
\newblock E -- a brainiac theorem prover.
\newblock \emph{AI Communications}, 15\penalty0 (2):\penalty0 111--126, 2002.

\bibitem[Schulz(2013)]{Schulz13}
S.~Schulz.
\newblock System description: {E} 1.8.
\newblock In K.~L. McMillan, A.~Middeldorp, and A.~Voronkov, editors,
  \emph{LPAR}, volume 8312 of \emph{LNCS}, pages 735--743. Springer, 2013.
\newblock ISBN 978-3-642-45220-8.
\newblock \doi{10.1007/978-3-642-45221-5_49}.
\newblock URL \url{http://dx.doi.org/10.1007/978-3-642-45221-5_49}.

\bibitem[Sutcliffe(2013)]{sutcliffe2013-j6}
G.~Sutcliffe.
\newblock The 6th {IJCAR} automated theorem proving system competition -
  {CASC-J6}.
\newblock \emph{{AI} Commun.}, 26\penalty0 (2):\penalty0 211--223, 2013.
\newblock \doi{10.3233/AIC-130550}.
\newblock URL \url{http://dx.doi.org/10.3233/AIC-130550}.

\bibitem[Urban(2004)]{Urb04-MPTP0}
J.~Urban.
\newblock {MPTP - Motivation, Implementation, First Experiments}.
\newblock \emph{J. Autom. Reasoning}, 33\penalty0 (3-4):\penalty0 319--339,
  2004.
\newblock \doi{10.1007/s10817-004-6245-1}.

\bibitem[Urban(2006)]{Urban06}
J.~Urban.
\newblock {MPTP} 0.2: Design, implementation, and initial experiments.
\newblock \emph{J. Autom. Reasoning}, 37\penalty0 (1-2):\penalty0 21--43, 2006.

\bibitem[Urban(2015)]{blistr}
J.~Urban.
\newblock {BliStr: The Blind Strategymaker}.
\newblock In G.~Gottlob, G.~Sutcliffe, and A.~Voronkov, editors, \emph{GCAI
  2015. Global Conference on Artificial Intelligence}, volume~36 of \emph{EPiC
  Series in Computing}, pages 312--319. EasyChair, 2015.

\bibitem[Urban et~al.(2008)Urban, Sutcliffe, Pudl{\'a}k, and
  Vysko\v{c}il]{US+08-long}
J.~Urban, G.~Sutcliffe, P.~Pudl{\'a}k, and J.~Vysko\v{c}il.
\newblock {MaLARea SG1 - Machine Learner for Automated Reasoning with Semantic
  Guidance}.
\newblock In A.~Armando, P.~Baumgartner, and G.~Dowek, editors, \emph{IJCAR},
  volume 5195 of \emph{LNCS}, pages 441--456. Springer, 2008.
\newblock ISBN 978-3-540-71069-1.

\bibitem[Wang et~al.(2016)Wang, Hutter, Zoghi, Matheson, and
  de~Feitas]{wang2016bayesian}
Z.~Wang, F.~Hutter, M.~Zoghi, D.~Matheson, and N.~de~Feitas.
\newblock Bayesian optimization in a billion dimensions via random embeddings.
\newblock \emph{Journal of Artificial Intelligence Research}, 55:\penalty0
  361--387, 2016.

\bibitem[Zhang and Shasha(1989)]{Zhang:1989:SFA:76071.76082}
K.~Zhang and D.~Shasha.
\newblock Simple fast algorithms for the editing distance between trees and
  related problems.
\newblock \emph{SIAM J. Comput.}, 18\penalty0 (6):\penalty0 1245--1262, Dec.
  1989.
\newblock ISSN 0097-5397.
\newblock \doi{10.1137/0218082}.
\newblock URL \url{http://dx.doi.org/10.1137/0218082}.

\end{thebibliography}


\end{document}